\documentclass[aps,prl,twocolumn,footinbib,superscriptaddress,longbibliography,amsmath,amssymb]{revtex4-1}
\usepackage[colorlinks,citecolor=blue,linkcolor=red]{hyperref}
\usepackage{graphicx}
\usepackage{subfigure}
\usepackage{epsfig}
\usepackage{dcolumn}
\usepackage{bm}
\usepackage{times}
\usepackage{amsmath}
\usepackage{float}
\usepackage{multirow}
\usepackage[normalem]{ulem}
\usepackage{color}
\usepackage{pifont}

\begin{document}

\title{Twisted Heating-Cooling Transition of Near-field Radiation in Drifted Metasurfaces}

\author{Jiebin Peng}
\email{Corresponding Email: jiebin.peng@gdut.edu.cn}
\affiliation{Center for Phononics and Thermal Energy Science, China-EU Joint Lab on
Nanophononics, Shanghai Key Laboratory of Special Artificial Microstructure Materials and
Technology, School of Physics Science and Engineering, Tongji University, 200092 Shanghai,
China}
\affiliation{School of Physics and Optoelectronic Engineering, Guangdong University of Technology, Guangzhou 510006, Guangdong Province, PR China}

\author{Zi Wang}
\affiliation{Center for Phononics and Thermal Energy Science, China-EU Joint Lab on
Nanophononics, Shanghai Key Laboratory of Special Artificial Microstructure Materials and
Technology, School of Physics Science and Engineering, Tongji University, 200092 Shanghai,
China}

\author{Jie Ren}
\email{Corresponding Email: xonics@tongji.edu.cn}
\affiliation{Center for Phononics and Thermal Energy Science, China-EU Joint Lab on
Nanophononics, Shanghai Key Laboratory of Special Artificial Microstructure Materials and
Technology, School of Physics Science and Engineering, Tongji University, 200092 Shanghai,
China}

\begin{abstract}
The magic angle twisted bilayer systems give rise to many exotic phenomena in two-dimensional electronic or photonic platforms. Here, we study the twisted near-field energy radiation between graphene metasurfaces with nonequilibrium drifted Dirac electrons. Anomalously, we find unconventional radiative flux that directs heat from cold to hot. This far-from-equilibrium phenomenon leads to a heating-cooling transition beyond a thermal magic twist angle, facilitated by twist-induced photonic topological transitions. The underlying mechanism is related to the spectrum match and mismatch caused by the cooperation between the non-reciprocal nature of drifted plasmon polaritons and their topological features. Furthermore, we report the unintuitive distance dependence of radiative energy flux under large twist angles. The near-field radiation becomes thermal insulating when increasing to a critical distance, and subsequently reverses the radiative flux to increase the cooling power as the distance increases further. Our results indicate the promising future of nonequilibrium drifted and twisted devices and pave the way towards tunable radiative thermal management.
\end{abstract}

\maketitle

The manipulation of thermal-photon carrier density in the near-field has been extensively studied for many technological applications, including enhanced radiative cooling\cite{Guha2012,YaoZhai2017}, thermophotovoltaic\cite{Messina-2013,Lenert2014}, and near-field imaging\cite{Kittel2005}. Inspired by the concept of magic-angle twisted bilayer graphene\cite{Yuan-2018,Sunku-2018}, a magic-angle twisted photonic hyperbolic metasurface, such as a periodic array of graphene ribbons\cite{Dai-2015,Guangwei-2020}, can be used to break the geometric symmetry and induce a topological transition in the isofrequency dispersion to control the photon carrier density\cite{Luojunscience2023}. Additionally, there are similar methods to control near-field heat transfer, such as anisotropic and gaint thermal magnetoresistance\cite{Latella-2017,AbrahamEkeroth2018}, twist-induced radiative thermal switches\cite{Fan-2020,He:20,Gaomin-2021,Jiebin-2021,Zhou2021}, and twisted thermophotovoltaic systems\cite{Tuning2023}. However, most twisted methods are carried out with passive systems and rely on the local equilibrium approximation, which is ultimately limited by the temperature bias that restricts the direction of energy flow always from hot to cold.

Recently, various methods have been explored to overcome the limitations of temperature bias and achieve active cooling effects. These methods include spatiotemporal modulation of the material permittivity\cite{Buddhiraju2020,Renwen2023}, rotation of the nanostructures\cite{Control2023}, and regulation of photonic chemical potential\cite{Henry1996,Chen2015,Zhu2019} or current-biased graphene\cite{Tang2021PRL}. Owing to nonequilibrium nature of current-biased graphene, the distribution of radiation photons in current-biased graphene metasurface follows a fluid-like behavior, exhibiting non-reciprocity with time-reversal symmetry breaking\cite{NLD2017,natureDong2021,natureZhao2021}. The non-reciprocal hyperbolic plasmon polaritons in current-biased graphene metasurface thus emerge as an ideal platform for exploring the possibilities of twist-tunable heating-cooling transitions, resulting in a thermal magic angle.

In this Letter, we study the near-field radiative energy transfer between two hyperbolic graphene metasurfaces, with a focus on breaking both time-reversal and geometric symmetry. Specifically, we apply nonequilibrium drift current and twist to one of the graphene metasurfaces and explore the cooperative interplay between the non-reciprocal nature of drifted plasmon polaritons and the twist-induced topological transition. Our findings reveal the presence of a transition angle between heating and cooling, as well as cooling flux beyond the magic twist angle. Notably, we observe an anomalous distance effect for large twist angles, which originates from the distinct distance dependencies of the heating and cooling modes. These results advance our understanding of near-field energy radiation and would have implications for the future development of thermal management.

\begin{figure*}
  \centering
  \includegraphics[width=1\textwidth]{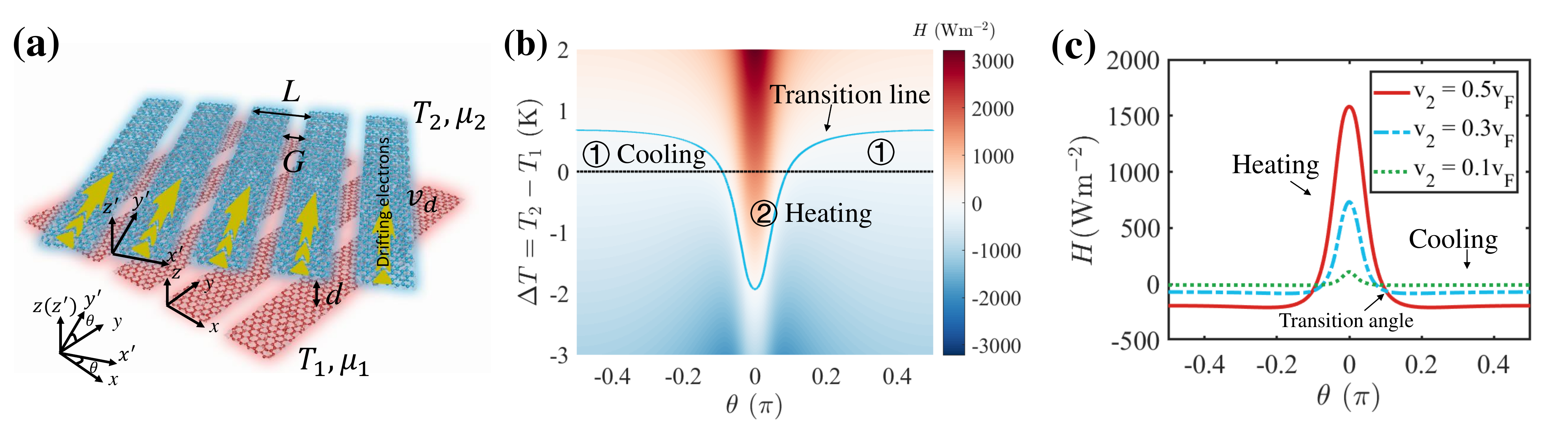}
  \caption{(a) A schematic setup for twisted and drifted radiative energy transfer between two graphene metasurfaces with a gap separation $d$. The top layer is driven with an external voltage and has drifted electrons with constant velocity $v_d$. $y'$ is defined as the direction of drift velocity in the top layer and tunable twist angle $\theta$ is defined as the anticlockwise rotation of $y'$ with respect to $y$ axis of the bottom layer. The chemical potentials of two layers are equal ($\mu_1=\mu_2=0.1 \mathrm{eV}$). (b) Color plot for the energy flux as a function of twist angle $\theta$ and temperature bias $\Delta T=T_2-T_1$ with fixed $T_1$. The blue-solid line is the transition line with zero energy flux. (c) Energy flux as a function of twist angle $\theta$ with different drift velocities in zero temperature bias condition ($\Delta T = 0$). The magic angle is defined as the transition angle between positive and negative energy flux. $v_F$ is the fermi velocity of graphene. The gap separation $d$ is 100 nm and $T_1=300 \mathrm{K}$.}
  \label{Fig1}
\end{figure*}

\emph{Near-field radiative energy transfer.--}We consider near-field radiative energy transfer between passive- and active- graphene metasurface with temperature differences $T_{1(2)}$ and twist angle $\theta$ [See Fig.~\ref{Fig1}(a)]. The bottom graphene layer is a passive layer without drift electrons($v_1=0$) and the top graphene layer is an active layer with drift electrons ($v_2 = v_d$) in our calculation. A Cartesian coordinate system $xyz$ ($x'y'z'$) is defined at the bottom (top) layer and the $y$($y'$) axis is along the direction of drift electron velocity. The twist angle $\theta$ is defined as the angle between the $y'$ and $y$ axis.  From fluctuational electrodynamics\cite{Polder-1971,Volokitin-2007}, the radiative energy flux can be expressed as
\begin{align} \label{Energy_flux}
   H=\int_0^{\infty} \frac{d\omega}{2\pi}  \int_{-\infty}^{\infty} \frac{d{\bf{k}}}{4\pi^2} \hbar\omega (n_2-n_1) {\cal{Z}}(v_d,\theta),
\end{align}
where ${\bf{k}}=[k_x,k_y]$ is the in-plane wave vector. $n_{1(2)}=1\big/ [e^{\hbar (\omega - k_yv_{1(2)})/k_b T_{1(2)}} - 1]$ is the drift Bose distribution with drift velocity $v_{1(2)}$ in $y$ direction and $T_{1(2)}$ is temperature of bottom(top) layer\cite{Henry1996,Volokitin2008,Volokitin2011}. For convenience, we define the energy transmission function of the energy flux in Eq. \ref{Energy_flux} as:
\begin{align}
      & {\cal{F}}(\omega,k_x,k_y) = \hbar\omega (n_2-n_1) {\cal{Z}}(v_d,\theta).
\end{align}
The energy flux spectrum $h(\omega)$ is integrand function of $H$ in frequency space, i.e., $H=\int_0^{\infty} h(\omega) d\omega /2\pi$. The properties of energy transmission function in $\omega-k_y$ panel (${\cal{F}}_{\omega k_y}$) can be described by the function of $\cal{F}$ after integrating  $k_x$, i.e., ${\cal{F}}_{\omega k_y} = \int {\cal{F}} dk_x$. The energy transmission coefficient ${\cal Z}(\omega,k_x,k_y)$ with twist angle $\theta$ reads
\begin{align}
{\cal Z }= \left\{
\begin{aligned}
	&{{\rm Tr}[({\bf I}-{\bf R}^\dagger_2{\bf R}_{2}) {\bf D} ({\bf I}-{\bf R}_1{\bf R}^\dagger_1) {\bf
			D}^\dagger] }, & |{\bf{k}}| < k_0 , \\
	&{{\rm Tr}[({\bf R}^\dagger_2 -{\bf R}_2) {\bf D} ({\bf R}_1 - {\bf R}^\dagger_1) {\bf D}^\dagger] }
	e^{-2|k_{z}|d}, &  |{\bf{k}}| > k_0,
	\end{aligned}
\right.
\end{align}
where $k_0=\omega/c$ is the wave vector in vacuum, $c$ is light velocity in vacuum and $k_z= \sqrt{(\omega/c)^2-{\bf{k}}^2}$ is the  in(out)-of-plane wave vector in vacuum. ${\bf I}$ denotes the identity matrix. The Fabry-Perot-like denominator matrix is written as ${\bf D}=({\bf I}-{\bf R}_1 {\bf R}_2 e^{2ik_z d})^{-1}$. $ {\bf R}_{1(2)} $ is the reflection coefficient matrix and related to the optical conductivity tensor of the bottom(top) metasurface.

\emph{Twisted optical conductivity tensor of the graphene metasurface with drift electrons.--} Without loss of generality, we take the example of graphene metasurface. The graphene strip periodicity has width $L$ and the air gap $G$ is the separation distance between the neighboring graphene strips. The deeply subwavelength periodicity assumption is satisfied in our calculation, i.e., $L = 10 {\rm nm} \ll \lambda_{tm}$ (where $\lambda_{tm}$ is the room temperature thermal wavelength). So that optical conductivity tensor $\overline{\sigma}^{\rm eff}= \left[ \sigma_{xx}^{\rm eff} , \sigma_{xy}^{\rm eff} ; \sigma_{yx}^{\rm eff} , \sigma_{yy}^{\rm eff} \right]$ of graphene metasurface can be characterized with the effective medium theory\cite{Gomez-Diaz2015}:
\begin{align}
\begin{split}
  \sigma^{\rm eff}_{xx} =& \frac{L\sigma_{xx}^g \sigma_{xx}^C}{W\sigma_{xx}^C+(L-W)\sigma_{xx}^g}, \sigma^{\rm eff}_{xy(yx)} = \sigma^{\rm eff}_{xx} \frac{W\sigma_{xy}^g}{L\sigma_{xx}^g},\\
  &\sigma^{\rm eff}_{yy} = \frac{W}{L}\sigma_{yy}^g-\frac{W\sigma_{yx}^g\sigma_{xy}^g}{L\sigma_{xx}^g} +\frac{\sigma^{\rm eff}_{yx} \sigma^{\rm eff}_{xy}}{\sigma^{\rm eff}_{xx}},
\end{split}
\end{align}
where $\sigma_{xx}^C = -i\frac{\omega \epsilon_0 L}{\pi} {\rm In} \left[  {\rm csc} \left( \frac{\pi }{2} \frac{L-W}{L} \right) \right]$ is the nonlocal correction parameter taking into account the near-field coupling of adjacent ribbons.
$\overline{\sigma}^g= \left[ \sigma_{xx}^g , \sigma_{xy}^g ; \sigma_{yx}^g , \sigma_{yy}^g \right]$ is the nonlocal optical conductivity tensor of graphene \cite{Lovat2013}. Furthermore, we use a Doppler shift model~\cite{NLD2017} to study the drift effects of the metasurface in $y$ direction:
\begin{align}\label{Drift_oc}
  \overline{\sigma}_{d}(\omega,v_d) \approx \frac{\omega}{\omega - k_y v_d} \overline{\sigma}_{\rm eff}(\omega-k_y v_d).
\end{align}
The optical properties of graphene metasurface are anisotropic and the corresponding surface plasmon-polaritons becomes non-reciprocal due to the drifted electrons along the strips. The general formula for the optical conductivity tensor in $k_x$-$k_y$ space with twist angle $\theta$ can be written as\cite{Correas-Serrano2015}:
\begin{align}\label{Rotation}
  \overline{\sigma}(\theta) =&   {\cal{R}}(\theta)
   \frac{1}{k_\rho}
  \begin{bmatrix}
  k_x & -k_y \\
  k_y & k_x
  \end{bmatrix}
  \overline{\sigma}_{d}
  \frac{1}{k_\rho}
  \begin{bmatrix}
  k_x & k_y \\
  -k_y & k_x
  \end{bmatrix}
    {\cal{R}}^{-1}(\theta),
\end{align}
where ${\cal{R}}(\theta)$ is the 2D rotation matrix and the positive $\theta$ corresponds to anticlockwise rotation of the top active graphene metasurface.

During the numerical calculation, the temperature of bottom passive layer $T_1$ is fixed at room temperature ($\sim 300\mathrm{K}$). The chemical potentials of two layers are equal ($\sim 0.1\mathrm{eV}$). The strip periodicity $L$ for graphene metasurface is set as $10$ nm and the air gap $G$ is set as 5 nm. The relaxation time of graphene is $3.6 \times 10^{-13}$s and the Fermi velocity $v_F$ takes $10^6 \mathrm{ms^{-1}}$. The cut-off wave vector in the ${\bf{k}}$-integration is set as 500$k_0$.

\emph{Thermal magic twist angle of heating-cooling transition.--} Fig. \ref{Fig1}(b) shows the energy flux as a function of twist angle $\theta$ ranging from $-0.5\pi$ to $0.5\pi$  and temperature bias $\Delta T=T_2-T_1$ with fixed $T_1=300\mathrm{K}$. We observe the active heating and cooling behaviors in ($\theta -\Delta T $) phase diagram, that is the positive energy flux against the temperature gradient. The heating region \ding{173} is found to be close to zero twist angle, where the active-cold layer $2$ can counterintuitively transfer energy into the passive-hot layer $1$. By rotating the active layer beyond the transition twist angle ($\approx \pm0.1 \pi$), we can transform from the heating region \ding{173} into the cooling region \ding{172}, where the active-hot layer $2$ can counterintuitively absorb energy from the passive-cold layer $1$. Fig. \ref{Fig1}(c) reveals the drift velocity $v_d$ dependence of the heating-cooling transition behavior with zero temperature differences: the heating peak is more pronounced at elevated $v_d$ near to zero twist angle; however, the cooling power is unusually robust at large twist angle $\theta$. Further investigation implies that the resulting transition is related to the spectrum match and mismatch caused by the cooperation between the non-reciprocal nature of drifted plasmon polaritons and their topological features near the magic twist angle. Without loss of generality, we fix the temperature difference at zero in the following discussions.

\begin{figure}
  \centering
  \includegraphics[width=0.45\textwidth]{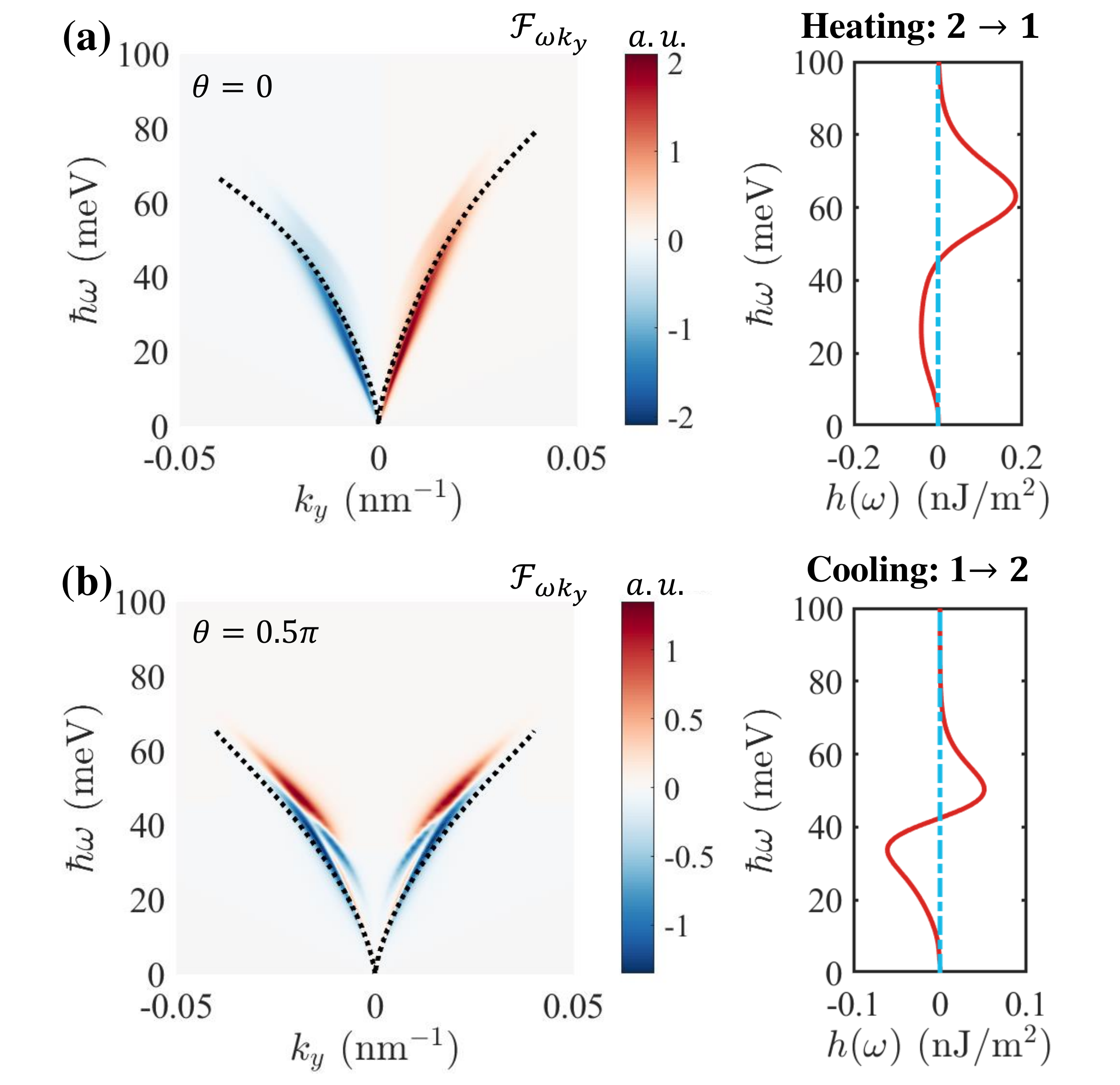}
  \caption{(a) Left: color plot for energy transmission function (${\cal{F}}_{\omega k_y}$) as a function of ($k_y,\omega$) with $\theta=0$. The dot-line is dispersion of surface plasmon polariton at $\omega-k_y$ panel in the Voigt configuration with $k_x = 0$. Right: the corresponding spectrum function $h(\omega)$ with $\Delta T=0$. (b) Similar as (a) except for $\theta=0.5 \pi$. The dot-line in the left is the dispersion of the surface plasmon polaritons with $k_x = 150 k_0$. The drift velocity $v_d = 0.5 v_F$ and the gap separation $d = 100 $nm. a.u., arbitrary units}
  \label{Fig2}
\end{figure}

We first discuss the dispersion of the nonreciprocal surface plasmon polaritons in the $\omega-k_y$ panel and its evolution with twist angle. The dispersion for the nonreciprocal surface plasmon polaritons in the cavity formed by the passive- and active- graphene metasurface can be obtained as the singularity of the Fabry-Perot-like denominator matrix:
\begin{align}\label{EQ_Dis}
    \mathrm{Tr} \left( {\bf I}-{\bf R}_1 {\bf R}_2 e^{-2|k_z| d} \right) =0,
\end{align}
 The symmetric breaking in the surface plasmon polariton dispersion ($\omega(k_y) \ne \omega(-k_y)$) emerges at $\theta = 0$ as shown in Fig. \ref{Fig2}(a): such nonreciprocal effects are caused by the nonequilibrium drift and increase with the velocity $v_d$. The roles of such nonreciprocal behavior in the radiative energy transfer are illustrated in terms of the color plot of the $k_x$-integrated energy transmission function ${\cal{F}}_{\omega k_y}$ which maintains a good agreement with the dispersion and presents the asymmetric distributions of the cooling (absorbing energy from layer $1$ to $2$) and heating (transferring energy from layer $2$ to $1$) modes in the negative and positive $k_y$, respectively. The direction of net energy flux depends on the competition  between these cooling and heating modes. Stating from the low frequency region at the corresponding spectrum function $h(\omega)$ at $\theta = 0$, the cooling modes are dominated channels at $0<\omega<45 \mathrm{meV}$, where the nonreciprocal effects are not noticeable. With the increased frequency, the heating modes become the dominant channels due to the increased nonreciprocal effects. This enables a positive net energy flux at $\theta = 0$.

However, the competitive situation between the cooling and heating modes can be reversed by rotating the active layer to large angles. Fig. \ref{Fig2}(b) shows the recovered symmetric dispersion of the surface plasmon polaritons in $\omega-k_y$ panel with fixed $k_x=150 k_0$ and we can observe the negative net energy flux at $\theta = 0.5 \pi$. To understand this, we show a similar symmetric recovery of the $k_x$-integrated energy transmission function ${\cal{F}}_{\omega k_y}$. The envelope curve of ${\cal{F}}_{\omega k_y}$ also maintains good agreement with the dispersion line, but the distributions of the cooling and heating modes in the $\omega-k_y$ panel are different from those shown in Fig. \ref{Fig2}(a): two bands for heating modes in the high frequency region and four bands for cooling modes in the low frequency region, which is analogous to the band splitting effects in electronic band structures. Quantitatively, the extreme anisotropic nature of the bottom passive layer leads to different drag coefficients at different twist angles via near-field coupling to the drifted Dirac electrons in the top active layer, resulting in a redistribution of the cooling and heating modes in momentum $k$-space. The corresponding spectrum function $h(\omega)$ shows that compared with the spectrum function in Fig. \ref{Fig2}(a), the peak for higher-frequency heating modes decreases rapidly but the peak for lower-frequency cooling modes increases at $\theta= 0.5 \pi$. Therefore,  the interplay of the nonreciprocal and anisotropic Fizeau drag effects gives rise to a negative net energy flux after twisting the active layer at large twist angles.
\begin{figure*}
  \centering
  \includegraphics[width=1\textwidth]{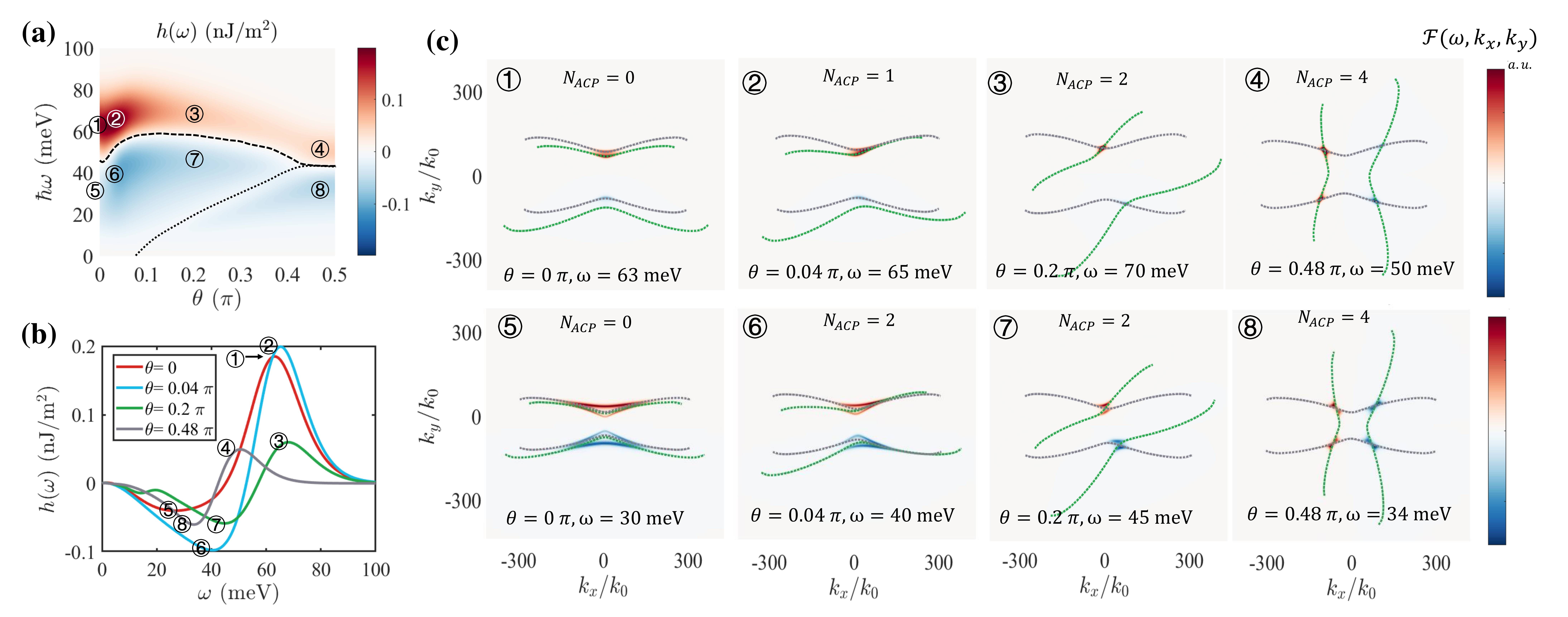}
  \caption{(a) Color plot for spectrum function $h(\omega)$ as a function of ($\theta,\omega$): the transition line between heating and cooling modes (dashed line),  the transition line between cooling modes with different topological features (dotted line), the transition angles for the creation and destruction of different topological polaritons (dash-dotted line). \ding{172}-\ding{175} are the resonant frequency regions for heating peaks at $\theta = 0, 0.04 \pi, 0.2 \pi, 0.48 \pi$ and \ding{176}-\ding{179} are similar regions for cooling peaks. (b) Spectrum function at $\theta = 0, 0.04 \pi, 0.2 \pi, 0.48 \pi$. (c) Energy transmission function ${\cal{F}}(\omega,k_x,k_y)$ at different resonant frequency regions. The grey(green)-dotted lines are the dispersion lines of surface plasmon polaritons in single passive(active) layer. The gap separation $d$ is 100nm and temperature difference is set as zero. a.u., arbitrary units}
  \label{Fig3}
\end{figure*}

\emph{Twist-induced topological transition in nonreciprocal isofrequency spectrum.--} We then study the $\theta$-dependence of the spectrum function $h(\omega)$ and energy transmission function $ {\cal{F}}(\omega,k_x,k_y)$ at resonant frequencies of cooling and heating modes. The color plot of $h(\omega)$ is divided into several parts with different topological features at Fig. \ref{Fig3}(a). The circled numbers in Fig. \ref{Fig3}(a-b) correspond to resonant frequency regions in $\theta-\omega$  phase space and spectrum functions, at $\theta=0, 0.04\pi, 0.2\pi, 0.48\pi$, respectively. We can find that the resonant heating and cooling modes both have blue shift in the regime with twist angle $0<\theta<0.1\pi$ and red shift in the regime with twist angle $0.2\pi<\theta<0.5\pi$.  Note that when $0.1\pi<\theta<0.2\pi$, the resonant peaks of both modes are significantly reduced, but the reduction of heating modes is severer than that of cooling modes due to different topological transitions. We use the number of anti-crossing points ($N_{ACP}$) of the dispersion lines of each isolated layer to describe the topological quantity\cite{Guangwei-2020}. The dispersion lines of surface plasmon polaritons supported by a single graphene metasurface can be calculated by the poles of the reflection coefficient matrix:
\begin{align}
    (2 c\epsilon_0 k_z/k_0 + \sigma_{yy}^{\rm eff})(2 c\epsilon_0  k_0/k_z + \sigma_{xx}^{\rm eff}) - \sigma_{xy}^{\rm eff} \sigma_{yx}^{\rm eff} =0,
\end{align}
where $\epsilon_0$ is the permittivity of vacuum.

Fig. \ref{Fig3}(c) shows that the energy transmission functions with heating and cooling bands (represented by the red and blue regions) maintain consistency with the dispersion line of the single graphene metasurface. However, they exhibit different topological transitions, with heating modes following $N_{ACP}=0\rightarrow 1 \rightarrow 2 \rightarrow 4$ and cooling modes following $N_{ACP}=0\rightarrow 2 \rightarrow 4$, as the twist angle $\theta$ increases. At small twist angles, heating and cooling modes take different topological transitions (i.e., $N_{ACP}=0\rightarrow 1 \rightarrow 2$ vs $N_{ACP}=0 \rightarrow 2$ from $\theta=0$ to $\theta=0.2\pi$), while demonstrating similar topological transitions at large twist angles (i.e., $N_{ACP}=2 \rightarrow 4$ from $\theta=0.2\pi$ to $\theta=0.4\pi$). Qualitatively, the sign of the spectrum will depend on the topological structure and photon density of states near the $N_{ACP}$ point in $k$-space. At $\theta=0$, $N_{ACP}$ equals zero due to the drift-induced nonreciprocal effects (\ding{172}\ding{176} in Fig. \ref{Fig3}(c)). Energy transfer channels can still be constructed via the scattering processes, but significant mismatch of heating or cooling bands in the high frequency heating modes is observed. As $\theta$ increases, the heating bands of heating modes form one nearly degenerate band in the $k_x-k_y$ space, while the cooling bands of heating modes remain large mismatch. Due to the large drift-induced mismatch of high frequency heating modes, this only results in a slight enhancement of the heating peak in the spectrum function (Fig. \ref{Fig3}(b)) when $N_{ACP}=0\rightarrow 1$. In contrast, the competition mechanism between heating and cooling bands of cooling modes greatly enhances the contributions of cooling modes. The cooling band of cooling modes with larger radius of curvature in high-$k$ region constructs more channels for energy transfer when $N_{ACP}=0\rightarrow 2$. We concluded that such manipulation of thermal-photon carrier density is related to the drift-induced non-reciprocity and twist-induced topological transitions. As the twist angle continues to increase, the flat degenerate bands split and undergo a hyperbolic-elliptic transition, similar to the photonic topological transition in ref\cite{Guangwei-2020}. However, all bands of heating and cooling modes undergo a similar topological transition ($N_{ACP}=2 \rightarrow 4$) at large twist angles, resulting in a saturated heat flux that no longer changes dramatically under further twisting.

\begin{figure}
  \centering
  \includegraphics[width=0.45\textwidth]{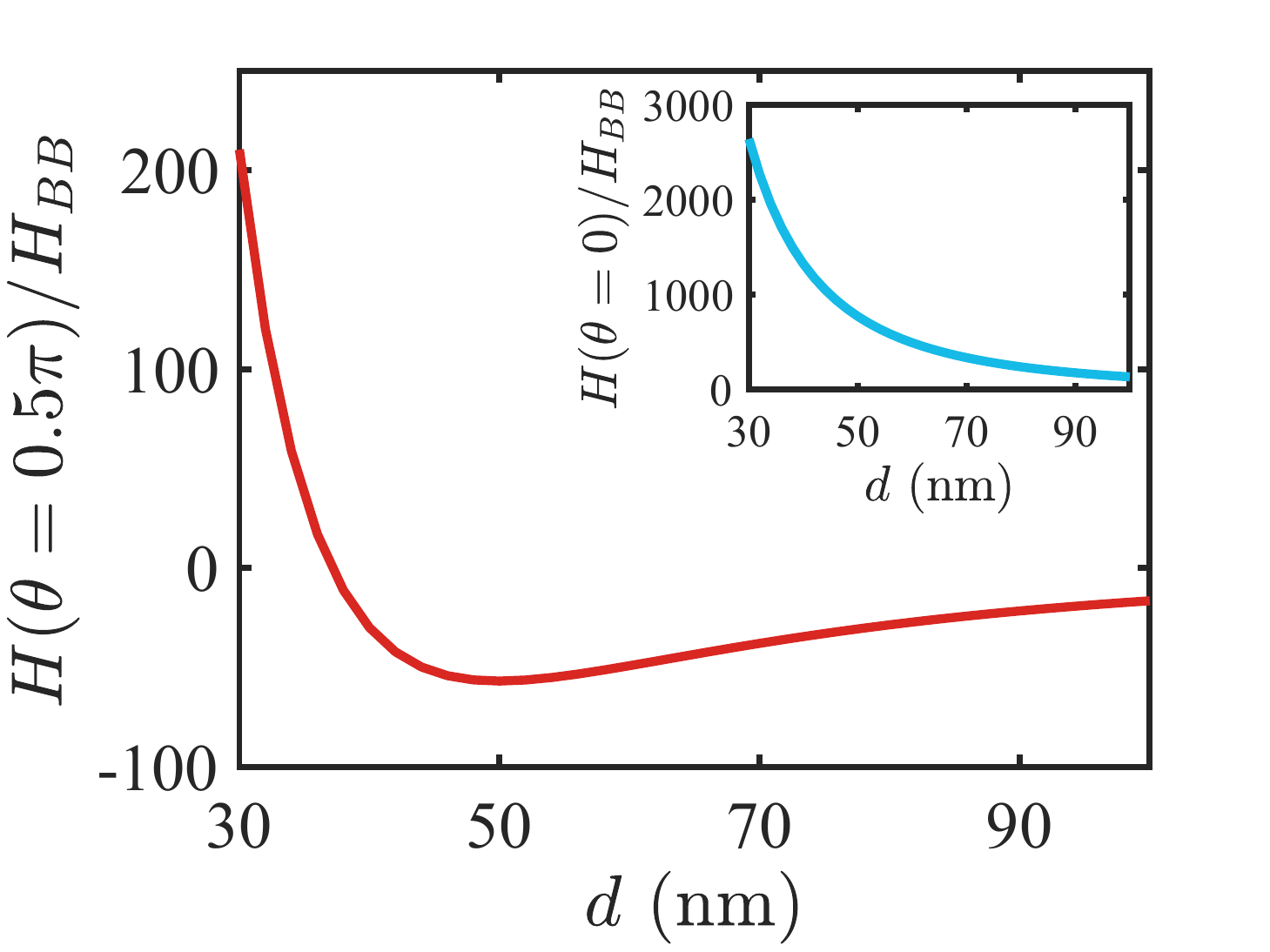}
  \caption{The ratio between the energy flux H and $H_{BB}$ as a function of gap separation $d$ with $\theta = 0.5 \pi$. The temperature difference $\Delta T= 0$ with $T_1 = 300$ K. $H_{BB} =\sigma T_{BB}^4$ is the radiosity of a blackbody at room temperature $T_{BB}=T_1$ and $\sigma$ is the Stefan-Boltzmann constant. The inset shows the ratio between energy flux with $\theta = 0$ and $H_{BB}$  versus gap separation $d$.}
  \label{Fig4}
\end{figure}
\emph{Abnormal distance effect at large twist angle.--} Another feature in Fig. \ref{Fig3}(c) is that the heating modes exist in the interval with higher $k$ value. Such heating modes take a greater attenuation coefficient in the vertical direction of plane compared with the cooling modes, which can lead to an abnormal distance effect. To verify these, we show the ratio between energy flux and blackbody limit versus the gap separation $d$ at $\theta=0.5\pi$ and $0$ in Fig. \ref{Fig4} (for convenience, we set a large-scale temperature difference for blackbody limit, i.e., $T_{BB}=300$K). The interval of gap separation is set as 30-100 nm to ensure the validity of effective medium theory applying on the metasurface of graphene strips. We find that there is nonmonotonic distance dependence at $\theta=0.5\pi$ (red line in Fig. \ref{Fig4}) and the energy flux can change the direction with gap separation d increasing. Similar phenomena are discussed in \cite{Tang2021PRL} by applying a large magnetic field in magneto-optic media, while here we take a rotation for the nonequilibrium active graphene metasurface. This is the predictable result of the interplay between nonreciprocal effects and photonic topological transitions: the heating mode with a larger attenuation coefficient will play the dominant role with a smaller gap separation. Notably, our calculations do not violate the second law of thermodynamics: the anomalous energy flow of the inverse temperature gradient can exist in the case of non-equilibrium drift.

In conclusion, we have studied the near-field energy transfer between magic-angle twisted graphene metasurfaces with drifted Dirac electrons. Our findings make a clear case that the current-drived graphene metasurfaces can possess a heating-cooling transition with a rotation regulation. Such behaviour is related to the nonreciprocal photon occupation number from the active graphene metasurface and hyperbolic anisotropy nature from the passive graphene metasurface. Under rotation regulation, the band splitting and degeneracy are observed at resonant frequencies. We therefore conclude that two types of photonic topological transitions exist for heating and cooling modes at a kind of thermal magic angle, arising from the interplay between nonreciprocal and non-local properties of graphene metasurface.
The calculations further reveal the unintuitive distance dependence of radiative energy flux under large twist angles. That is, the drifted and twisted near-field radiation becomes thermal insulating when increasing to a critical distance and subsequently reverses the radiative energy flux to anomalously increase the cooling power as increasing distance further.
Our work establishes a clear connection between microscopic photonic topological transition and macroscopic heating-cooling transition, which can pave a new way for nanoscale energy transport and thermal managements with nonreciprocity.

\begin{acknowledgments}
This work is supported by the National Natural Science Foundation of China (No. 11935010), the Shanghai Science and Technology Committee (Grants No. 23ZR1481200), and the Opening Project of Shanghai Key Laboratory of Special Artificial Microstructure Materials and Technology.
\end{acknowledgments}

\bibliography{Drift_ref}

\end{document}